\documentclass[10pt]{article}

\usepackage{amsmath}
\usepackage{amssymb}

\usepackage{graphicx}

\usepackage{cite}

\usepackage{color} 

\usepackage[labelfont=bf,labelsep=period,justification=raggedright]{caption}

\makeatletter
\renewcommand{\@biblabel}[1]{\quad#1.}
\makeatother

\date{\today}

\begin{document}

\begin{flushleft}
{\Large
\textbf{The missing assets and the size of Shadow Banking: an update}
}
\\
\bigskip
Davide Fiaschi$^{1}$, 
Imre Kondor$^{2,3}$, 
Matteo Marsili$^{4,\ast}$,
Valerio Volpati $^{5}$
\\
\bigskip
\bf{1} Dipartimento di Economia e Management, University of Pisa, Pisa, Italy
\\
\bf{2} Parmenides Foundation, Pullach b. M\"unchen, Germany
\\
\bf{3} London Mathematical Laboratory, London, UK
\\
\bf{4} The Abdus Salam International 
Centre for Theoretical Physics, Trieste, Italy
\\
\bf{5} International School for Advanced Studies (SISSA), 
Trieste, Italy
\\
$\ast$ E-mail: Corresponding author marsili@ictp.it
\end{flushleft}

\begin{abstract}
In a recent paper, using data from Forbes Global 2000, we have observed that the upper tail of the firm size distribution (by assets) falls off much faster than a Pareto distribution.
The {\em missing mass} was suggested as an indicator of the size of the Shadow Banking (SB) sector. 
This short note provides the latest figures of the missing assets for 2013, 2014 and 2015. In 2013 and 2014 the dynamics of the missing assets continued being strongly correlated with estimates of the size of the SB sector of the Financial Stability Board (FSB). In 2015 we find a sharp decrease in the size of missing assets, suggesting that the SB sector is deflating.
\end{abstract}

\section{Background}

\begin{figure}[h]
   \centering
  \includegraphics[width=3.4in]{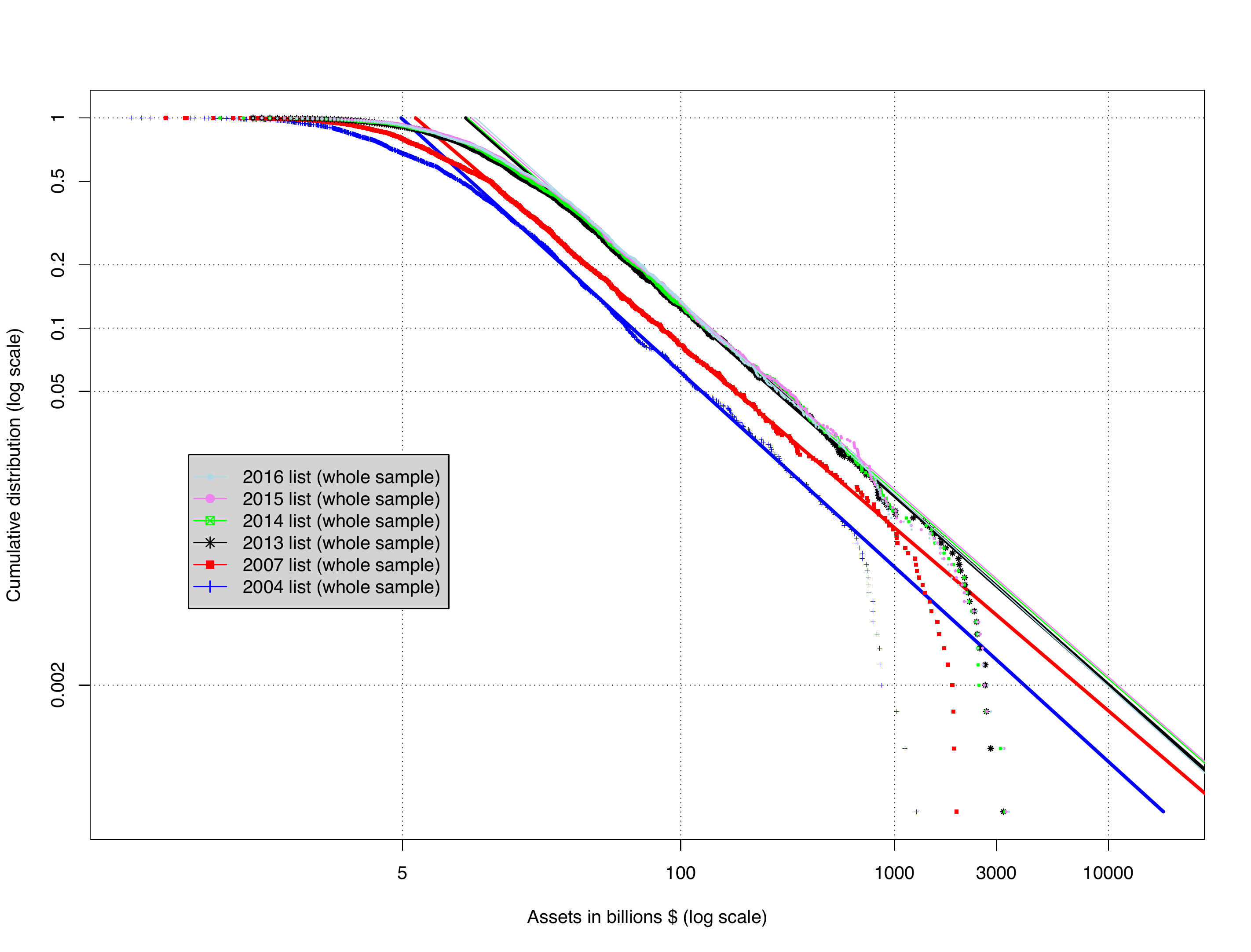}
    \includegraphics[width=2.6in]{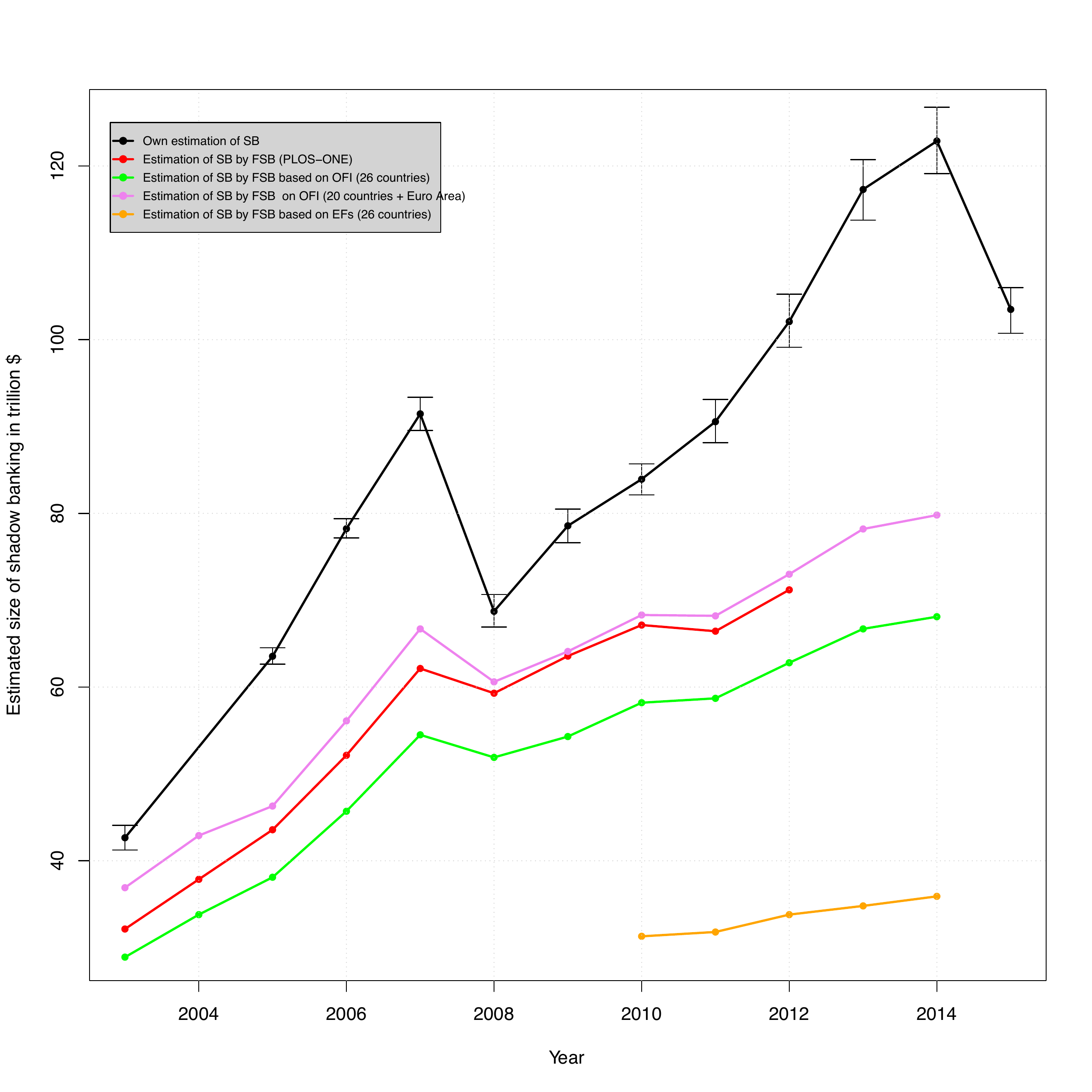}
   \caption{(Left) Cumulative asset size distribution of FG2000 firms in different years and the corresponding fits with a Pareto distribution $P(A>x)=cx^{-b}$. (Right) Comparison between the index $I_{SB}$ of FKMV, with the estimate of the size of SB made by FSB \cite{FSB} for the period 2003-2014. 
   The reported confidence bands for our estimate of SB are calculated on the basis of $\pm 2$ standard errors in the estimate of the coefficients of the power law distributions.}
   \label{fig:SBIndex_comparison}
\end{figure}

Taking the Forbes Global 2000 (FG2000) list as a snapshot of the global economy, Ref. \cite{PlosPaper} (hereafter referred to as FKMV) observed that the asset size distribution of global firms differs from the Pareto distribution predicted by proportional growth models. More specifically, the upper tail of the empirical distribution falls off much faster than a Pareto distribution, as shown in Fig. \ref{fig:SBIndex_comparison} (left). 

In order to compare the empirical distribution (i.e. the fraction of firms with assets $A>x$) with the theoretical Pareto distribution $P(A>x)=c x^{-b}$, we estimate the parameters $a=\log c$ and $b$ from linear fits in an intermediate range (see FKMV for details). The difference between the empirical and the theoretical distribution defines the {\em missing mass} $I_{SB}$, which is the total amount of assets that are missing from the upper tail of the distribution (see FKMV). The missing mass $I_{SB}$ is plotted in Fig. \ref{fig:SBIndex_comparison} (right) for the period 2004-2015. 

FKMV remark that: {\em i)} The largest firms in terms of asset size in the global economy are all in the finance sector, {\em ii)} the finance sector is expected to obey proportional growth dynamics and data corroborates this expectation. This expectation in turns predicts that the distribution of firms by asset size should be a Pareto distribution, {\em iii)} this prediction matches empirical data up to a cutoff (in the range of \$2-3 trillions) beyond which the distribution falls off very sharply. On this basis, FKMV argue that the missing mass can be taken as a measure of the amount of assets that would be missing with respect to the {\em ideal limit} of an unregulated economy dominated by proportional growth. Hence, FKMV suggested that the missing mass $I_{SB}$ could be used as a quantitative estimate of the size of the shadow banking (SB) sector (the Shadow Banking Index). 

Indeed, FKMV shows that $I_{SB}$ has been remarkably well correlated to estimates of the size of the  SB sector until 2012, although its numerical value is larger by approximately a factor of two. 

\section{Update}

After FKMV was published, the data for the Forbes Global 2000 list has been released for the years 2013, 2014 and 2015. This allows us to compute the missing mass $I_{SB}$ for these years as well. 
Fig. \ref{fig:SBIndex_comparison} shows that $I_{SB}$ has been remarkably consistent with the trend of the FSB estimates, in spite of the fact that the definition of SB and the way in which it has been measured has evolved in this period (see updated FSB estimates). 

The estimates of $I_{SB}$ with the 95\% confidence interval is reported in Table \ref{TableSBI}. The total assets and the ratio of the Shadow Banking sector to the total assets is also reported. Finally we report the estimated coefficients $a=\log c$ and $b$ of the fit of the Pareto distribution. 

In the 7 years after the 2007-2008 crisis, the SB sector has been growing unceasingly, in spite of all efforts that have been deployed to tame systemic risks in the financial system. Our analysis on the $2015$ FG2000 data shows that the missing mass $I_{SB}$ reached a peak in 2014 and decreased in 2015. Assuming, as in FKMV, that $I_{SB}$ is a measure of the size of the SB sector, this suggests that the inflationary trend in the SB sector has reverted or at least halted. 

We notice also that the total assets in 2015 have decreased with respect to the previous year, interrupting a growing trend that also persisted since the 2007-2008 crisis. Likewise the ratio of the missing mass $I_{SB}$ to the total assets has also sharply decreased in 2015.

\begin{table}[ht]
\centering
\begin{tabular}{rrrrrrr}
  \hline
Year & $I_{SB}$ & $95\%$ bound & Tot. assets ($TA$) & $I_{SB}/TA$ & $a$ & $b$ \\ 
  \hline
 2003 & 42.65 & [41.2, 44.1] & 68083.70 & 0.61 & 1.48 &  0.93 \\ 
   2005 & 63.54 & [62.6, 64.5] & 88490.54 & 0.71 & 1.51   &   0.89 \\ 
   2006 & 78.23 & [77.2, 79.4] & 102706.30 & 0.75 & 1.52 &   0.87 \\ 
   2007 & 91.45 & [89.5, 93.4] & 119395.27 & 0.75 & 1.61  &  0.86 \\ 
   2008 & 68.72 & [66.9, 70.7] & 124601.39 & 0.54 & 1.80  &  0.90 \\ 
   2009 & 78.56 & [76.6, 80.5] & 124024.78 & 0.62 & 1.78  &  0.89 \\ 
   2010 & 83.94 & [82.1, 85.7] & 138291.52 & 0.59 & 1.92  &  0.90 \\ 
   2011 & 90.56 & [88.1, 93.1] & 148848.00 & 0.59 & 2.03  &  0.90 \\ 
   2012 & 102.08 & [99.1, 105.2] & 158713.90 & 0.62 & 2.05  & 0.90 \\ 
   2013 & 117.30 & [113.8, 120.7] & 160971.30 & 0.71 & 2.06 & 0.89 \\ 
   2014 & 122.86 & [119.1, 126.8] & 164292.20 & 0.73 & 2.10 & 0.89 \\ 
   2015 & 103.49 & [100.7, 106.0] & 161461.85 & 0.62 & 2.19  &  0.91 \\ 
   \hline
   \end{tabular}
\caption{\label{TableSBI}Estimated shadow banking with 95\% confidence bands, total assets in the 2000 Forbes sample, ratio of the SBI and total assets. Estimates of the coefficients of the Pareto distribution $\log P(A>x)=a-b\log x$.}
\end{table}

We limit this short note to the crude reporting of the statistical analysis and leave any comment on possible relation to changes in the global financial system (regulation, QE, central clearing houses, etc) or in the global economy (e.g. China's slowdown) to a future contribution by us or by others.

\end{document}